\documentstyle{paper}
\input{epsf}
\begin{document}

\title{Green's functions for transmission of X-rays and $\gamma$-rays 
through cold media} 

\author{
Pawe{\l} MAGDZIARZ \\
{\it Astronomical Observatory, Jagiellonian University, Orla 171, 
30-244 Cracow, Poland}\\ 
Andrzej A. ZDZIARSKI\\
{\it Copernicus Astronomical Center, Bartycka 18, 00-716 Warsaw, Poland}\\
}
 
\maketitle

\section*{Abstract}

Using a Monte Carlo method, we study Compton scattering and absorption of
X-rays and $\gamma$-rays in cold media. We consider transmission of 
X/$\gamma$-rays through a shell of an arbitrary optical depth, for which 
we derive energy-dependent Green's functions. Fitting the Green functions
with simple analytical formulae is in progress.  
We also present a simple treatment of the effect of absorption
on Green's functions for Compton scattering, which allow to treat  
media with an arbitrary ionization state and chemical composition.
Our transmission Green's functions allow to treat Thomson-thick absorbers, 
e.g. molecular tori of Seyfert 2s. 

\section{Introduction} 

We consider transmission of X/$\gamma$-rays through a slab assuming incident 
photons perpendicular to the slab, and integrate outgoing photons over 
all angles. This is equivalent to transmission through a geometrically thin 
but optically thick shell when neglecting effect of reflection.  
We calculate energy-dependent Green's functions taking into account 
fully relativistic Compton scattering as well as bound-free absorption.  
Green's functions are calculated using a Monte Carlo method of White, 
Lightman and Zdziarski (1988). An analytical formula 
fits the Green functions for Compton scattering and factorize impact 
of absorption. 

\section{Photon transmission through Compton-thick media}

Fig.~1 presents the total Green functions and its decomposition on the
first three scattering orders for various energies of incident photons and
various Thomson thickness of the slab. For a low optical thickness, 
the second-order scattering is mostly due to photons propagating along
the slab, which makes the second order Green function relatively flat in 
the range of $\Delta y$=1-3 and the third order function peaked at
$\Delta y$=3 (cf. Fig.~1A). For a higher optical thickness, the second and
third order Green function is peaked in the forward and backward direction
due to photons propagating perpendicularly to the slab. The second and
third order function on Fig.~1C exhibits corresponding peaks near 
$\Delta y \gsim 0$ and $\Delta y \lsim 4$. 
For high energy incident photons, the Green 
functions are increasingly suppressed with increasing wavelength shift 
because the Klein-Nishina decline reduces the escape probability 
(cf. Fig.~1A and 1D). The high energy photons are moved quickly  
to lower energies by scattering, and thus attain diffusion limit after  
a first few scattering.  

\begin{figure}[t]
\centering
\begin{center}
\leavevmode
\epsfysize=18.5pc \epsfbox{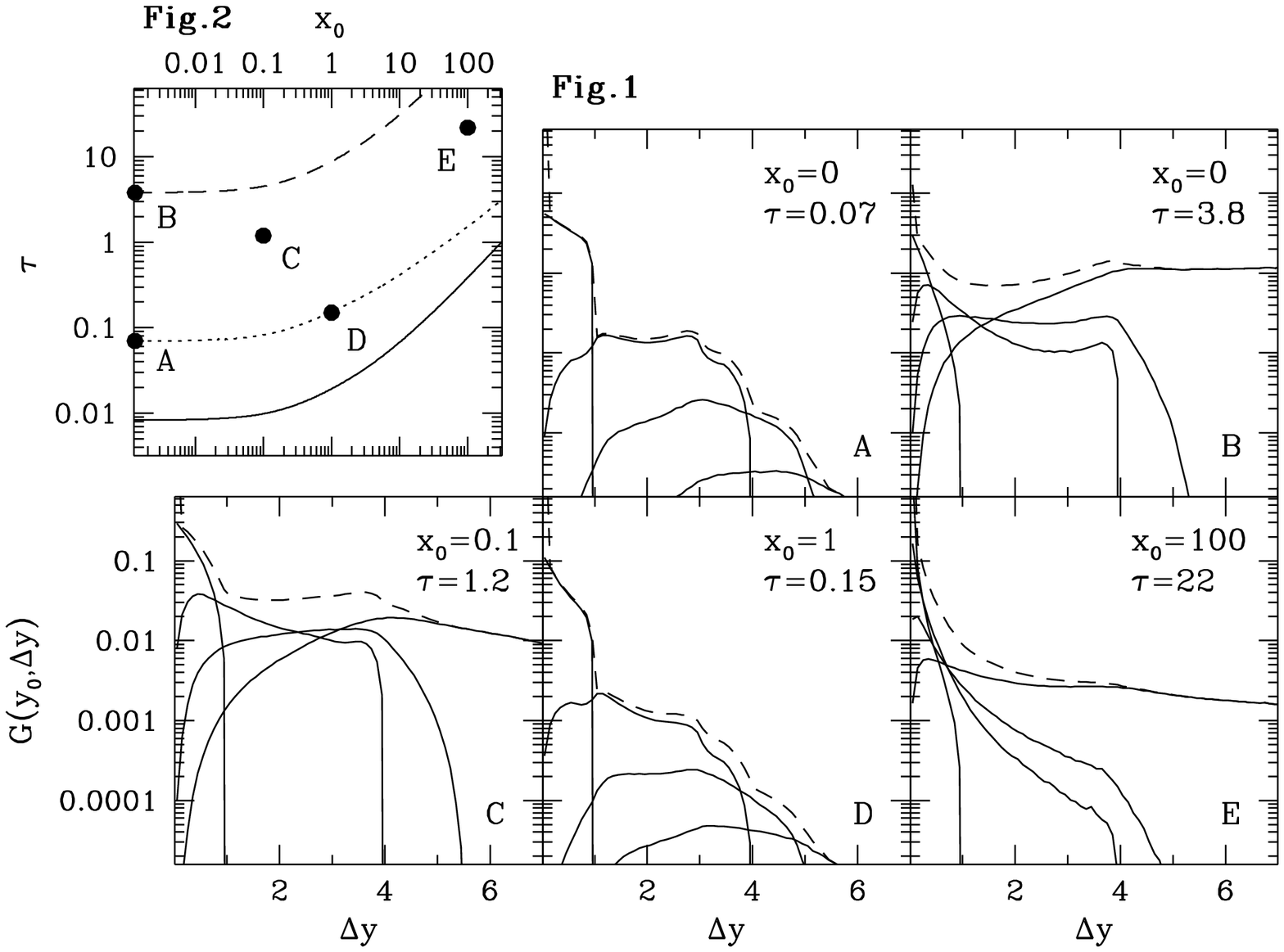}
\end{center}
   \caption{ The total Green functions (dashed line) in the points (A-E)
             marked on the Fig.~2. Solid lines mark contribution from
             subsequent orders Green's functions.}
   \caption{ Limits of the Thomson optical thickness for 1 per cent
             photons scattered by a geometrically thin shell (solid line), 
             and 1 per cent photons scattered more than ones (dotted line). 
             The dashed curve shows a limit of 1 per cent photons transmitted 
             without scattering. The area above this curve corresponds to 
             Compton scattering well-described by the diffusion 
             approximation. $x_0$ is incident photon energy in units of  
             electron rest mass.} 
\end{figure}
 
\section{The impact of absorption on Green's function} 
 
In solar composition matter, Compton scattering is accompanied by bound-free 
absorption, which gives important contribution up to energies of
about 40 keV, where Compton cross-section is significantly relativistic,
and no monochromatic approximation can be used. The Green function in  
the presence of absorption may be then written as:
\begin{equation} 
G_{abs}(y,\Delta y)=\sum_{i=1}^{\infty}G_i(y,\Delta y)
\prod_{k=1}^{i}\bar{\lambda}_k(y).
\end{equation}
Here $\bar{\lambda}_k$ is an effective single scattering albedo over the
distribution of photons after $k$ scattering. Since the average wavelengh 
shift of photon distribution in $k$-th order
appears at $\Delta y\approx k$ (relativistic effects shift the maximum to lower
values of $\Delta y$), a sufficient approximation is $\bar{\lambda}_k(y)
\approx\lambda(y+i-1)$. This allows us to factorize Green's function as:
\begin{equation}
G_{abs}(y,\Delta y)\approx G(y,\Delta y)
\exp\left(\int_0^{\Delta y-1}{\rm d}k\ln\left(
\lambda(y+k)\right)\right).
\end{equation}

\section{References}

\vspace{1pc}

\re
1. 	White, T. R., Lightman, A. P. and Zdziarski, A. A., 1988, {\it ApJ} {\bf 331}, 939.

\vspace{1pc}

\end{document}